# Topologically induced transparency in a two-phase metamaterial

Hafssaa Latioui[(1, 2)], Mário G. Silveirinha[(1,2)*]

[(1)] *Department of Electrical Engineering, University of Coimbra, and Instituto de Telecomunicações, 3030-290 Coimbra, Portugal*

[(2)]*University of Lisbon–Instituto Superior Técnico - Avenida Rovisco Pais, 1, 1049-001 Lisboa, Portugal*

## Abstract

It is theoretically and numerically demonstrated that a mixture of two topologically distinct material phases is characterized by an anomalous "transparency window" in a spectral range wherein the individual material phases are strongly reflecting. In particular, it is shown that a metamaterial formed by a metallic wire grid embedded in a magnetized plasma may support the propagation of waves with long wavelengths, notwithstanding the components, when taken separately, completely block the electromagnetic radiation. The effect is explained in terms of the topological properties of the magnetoplasmon. Furthermore, it is highlighted that some naturally available materials may be regarded as a mixture of two topologically distinct phases, and hence may be as well characterized by a similar anomalous transparency effect.

---

[*] To whom correspondence should be addressed: E-mail: *mario.silveirinha@co.it.pt*

# Main Text

Nonreciprocal effects in electromagnetics and nonreciprocal optical platforms have been recently the object of renewed interest and scrutiny, due to their unique and singular properties [1-12]. In particular, nonreciprocal systems enable one-way light flows and optical isolation [2-4, 8-12]. Nonreciprocal effects are typically obtained by tailoring the permittivity or permeability response with a static magnetic field, but alternative solutions have been explored recently [6-8, 10-11, 13-14].

Remarkably, some nonreciprocal systems have a topological nature [15-21]. Such systems are characterized by a topological integer that determines the number of topologically protected chiral edge states: the "Chern number". Indeed, the topology of a material can have remarkable consequences in the context of the electromagnetic propagation: when a topological material is paired with another material with a trivial topology, unidirectional scattering-immune gapless edge states emerge in a common bulk bandgap, a result known as the "the bulk edge correspondence" principle [18, 19, 21, 22]. Furthermore, it was recently shown that the photonic Chern number has a precise physical meaning: it is the *quantum* of the light-angular momentum spectral density in a photonic insulator cavity [22, 23].

Building on these previous works, here we theoretically demonstrate that topological edge modes (magneto-plasmons) may enable an anomalous "transparency effect" in a composite material formed by two topologically distinct fully-reflecting phases. We discuss how such an effect may be directly observed in naturally available materials and, in addition, we propose a realistic physical implementation relying on the metamaterial concept. It is assumed that the electromagnetic fields have a time dependence of the form $e^{-i\omega t}$.

As a starting point, let us consider a mixture of two materials, characterized by the permittivity tensors $\bar{\varepsilon}_1$ and $\bar{\varepsilon}_2$, respectively. The effective permittivity of the mixture depends on the volume fraction of each material, on the shape of the inclusions, lattice structure, etc [24, 25]. As an illustration, for now we use the simple mixing formula [26]:

$$\bar{\varepsilon} \approx (1-f_V)\bar{\varepsilon}_1 + f_V \bar{\varepsilon}_2. \qquad (1)$$

Here, $f_V$ is the volume fraction of the 2nd material. Furthermore, we shall focus on the case $f_V \ll 1$, so that $\bar{\varepsilon} \approx \bar{\varepsilon}_1 + f_V(\bar{\varepsilon}_2 - \mathbf{1})$. Then, if 1st material has a gyrotropic response and the 2nd material is isotropic with a Drude dispersion ($\varepsilon_2 = 1 - \omega_{p2}^2/\omega^2$), it follows that the composite medium is characterized by a dielectric function of the form $\bar{\varepsilon} = \varepsilon_t \mathbf{1}_t + \varepsilon_a \hat{\mathbf{z}} \otimes \hat{\mathbf{z}} + i\varepsilon_g \hat{\mathbf{z}} \times \mathbf{1}$ ($\mathbf{1}_t = \mathbf{1} - \hat{\mathbf{z}} \otimes \hat{\mathbf{z}}$) with

$$\varepsilon_t = 1 - \frac{\omega_p^2}{\omega^2 - \omega_c^2} - \frac{\tilde{\omega}_p^2}{\omega^2}, \qquad \varepsilon_g = \frac{1}{\omega}\frac{\omega_p^2 \omega_c}{\omega_c^2 - \omega^2}, \qquad (2)$$

and $\tilde{\omega}_p = \sqrt{f_V}\,\omega_{p2}$. The permittivity component $\varepsilon_a$ is not relevant in this study and hence is not shown. When $\tilde{\omega}_p = 0$ the permittivity dispersion is coincident with that of the 1st material, which is taken as a standard magnetized plasma (with bias magnetic field, $\mathbf{B}_0 = B_0 \hat{\mathbf{z}}$), e.g., it is similar to the response of magnetized semiconductors [27-28]. In such a context, $\omega_p$ is the plasma frequency of the electron gas and $\omega_c = -qB_0/m$ is the cyclotron frequency determined by the bias field ($q = -e$ is the negative charge of the electrons and $m$ is the effective mass) [29]. Magnetized plasmas and other gyrotropic media are generically topologically nontrivial materials [20, 21, 30-32]. On the other hand, the second material is characterized by a Drude dispersion model, ($\varepsilon_2 = 1 - \omega_{p2}^2/\omega^2$), which is reciprocal, and thus is topologically trivial. Hence, the global permittivity (2) models a composite formed by mixing two different topological phases, specifically a mixture of a topologically nontrivial material (1$^{st}$ phase) with a topologically trivial material (2$^{nd}$ phase).

The dispersion relation of an electric gyrotropic bulk medium for transverse magnetic (TM) polarized waves (wave propagation in the *xoy* plane with $H = H_z(x,y)\hat{\mathbf{z}}$) is given by $k^2 = \varepsilon_{ef}(\omega/c)^2$, with $\varepsilon_{ef} = (\varepsilon_t^2 - \varepsilon_g^2)/\varepsilon_t$ the equivalent permittivity of the gyrotropic material, $k^2 = k_x^2 + k_y^2$ and $k_x, k_y$ the wave vector Cartesian components. Figure 1a shows the band diagram of the TM polarized plane waves for a composite medium with $\omega_c/\omega_p = 0.5$ and $\omega_{p2} = \omega_p$, so that $\tilde{\omega}_p = \alpha \omega_p$ with $\alpha = \sqrt{f_V}$ determined by the volume fraction of the trivial material phase.

For $\alpha = 0$, the composite material has a single phase (1$^{st}$ component of the response) and the modes are organized in two branches. There are two photonic band-gaps highlighted as shaded regions in Fig. 1a. The respective gap Chern numbers are found as explained in Ref. [20] and are given in the insets. The Chern numbers are nonzero and hence the 1$^{st}$ phase is topologically non-trivial. On the other hand, the 2$^{nd}$ phase (with permittivity dispersion $\varepsilon_2 = 1 - \omega_{p2}^2/\omega^2$) has a topologically trivial low-frequency band-gap determined by $0 < \omega < \omega_{p2}$ (not shown).

Strikingly, for $\alpha = 0.5$ ($f_V = 0.25$) when the two different phases are mixed, a new low-frequency band ($0.10 < \omega/\omega_p < 0.21$) emerges well below the two relevant plasma frequencies ($\omega_p$, $\omega_{p2}$). This "transparency" window is the focus of our study. Figures 1c and 1d show how the transparency window changes under the variation of either $\omega_c/\omega_p$ or $\alpha$. Rather remarkably, the transparency window moves to even lower frequencies as either the volume fraction of the 2nd phase decreases ($\alpha$ decreases) or the bias magnetic field is reduced (lower values of $\omega_c/\omega_p$).

To demonstrate that the discovered low-frequency mode can be externally excited, next we study the wave scattering by a slab of the two-phase topological material. It is assumed that the material slab is surrounded by air. The air interfaces are at $y = 0$ and $y = d$, so that the propagation is along the $y$-direction. The magnetic field can be written as:

$$H_z = H_0^{inc} e^{ik_x x} \times \begin{cases} e^{-\gamma_0 y} + R e^{+\gamma_0 y} &, \quad y \leq 0 \\ A_1 e^{-\gamma_g y} + A_2 e^{+\gamma_g y} &, \quad 0 \leq y \leq d \\ T e^{-\gamma_0 (y-d)} &, \quad y \geq d \end{cases} \quad (3)$$

where $\gamma_0 = \sqrt{k_x^2 - (\omega/c)^2} = -i\sqrt{(\omega/c)^2 - k_x^2}$ and $\gamma_g = \sqrt{k_x^2 - \varepsilon_{ef}(\omega/c)^2}$, $R$ and $T$ are the reflection and transmission coefficients, $H_0^{inc}$ is the amplitude of the incident wave, $k_x = (\omega/c)\sin\theta_i$, and $\theta_i$ is the angle of incidence. The electric field in the gyrotropic material can be found using $\mathbf{E} = \dfrac{1}{-i\omega\varepsilon_0} \overline{\overline{\varepsilon}}^{-1} \cdot \left(\partial_y H_z \hat{\mathbf{x}} - \partial_x H_z \hat{\mathbf{y}}\right)$ with $\overline{\overline{\varepsilon}}^{-1} = \dfrac{1}{\varepsilon_{ef}}\left(\mathbf{1}_t - i\dfrac{\varepsilon_g}{\varepsilon_t}\hat{\mathbf{z}}\times\mathbf{1}_t\right) + \dfrac{1}{\varepsilon_a}\hat{\mathbf{z}}\otimes\hat{\mathbf{z}}$. Imposing the continuity of $E_x$ and $H_z$ at the interfaces one can find the transmission and reflection coefficients and the $A_1, A_2$ coefficients. In the following, we suppose that the dielectric function of the gyrotropic material is as in Eq. (2).

Figure 2a shows the amplitude of the transmission coefficient as function of the normalized thickness $L\omega_p/c$ for $\omega_c/\omega_p = 0.5$, $\alpha = 0.5$ and different values of incidence angle. The frequency of operation is $\omega/\omega_p = 0.15$ and thus lies roughly in the middle of the low-frequency transparency window (see Fig. 1a). As seen, the transmission coefficient exhibits a rather standard behavior with transmission peaks at the Fabry-Pérot resonances. Curiously, the transmission level improves for oblique incidence. Furthermore, Fig. 2b shows the transmission coefficient as function of the normalized frequency $\omega/\omega_p$ for different values of the normalized thickness $L\omega_p/c$. Interestingly, independent of the slab thickness, the wave can tunnel through the two-phase material near the frequency $\omega/\omega_p \approx 0.1$, which determines the lower edge of the transparency window in Fig. 1a. This property is reminiscent of the super-coupling effect characteristic of structures with near-zero refractive index [33, 34, 35]. As illustrated in Fig. 2c, the transmission characteristic is relatively robust to the effects of unavoidable material loss.

Next, we discuss how to physically realize the proposed structure using the metamaterial concept. We suggest that the two-phase topological material may be implemented relying on a standard magnetized electron gas as the host medium (e.g., a magnetized doped semiconductor), and a metallic wire grid. Evidently, the magnetized electron gas is expected to mimic the response of the nontrivial topological phase, whereas the wire grid is expected to imitate the trivial topological phase. The wire medium is formed by stacked 2D-metallic grids separated (along *z*) by the distance *a* [36, 37]. Each metallic grid consists of two perpendicular co-planar arrays of metallic strips oriented along the *x* and *y* directions with period *a*. The width of the strips is $w = 0.1a$. For simplicity, the metal is modeled as a perfect electric conductor (PEC). The grid design was based on the rough estimate that for $w = 0.1a$ the effective plasma frequency satisfies $\tilde{\omega}_p a / c \approx 2$, as if the host background were air [36, 38]. Thus, in order that $\tilde{\omega}_p = \alpha \omega_p$ one should use $a \sim \dfrac{2}{\alpha} \dfrac{c}{\omega_p} = \dfrac{1}{\pi \alpha} \lambda_{0,p}$, with $\lambda_{0,p}$ the free-space wavelength at the plasma frequency $\omega_p$. The host medium is taken as a lossy magnetized cold plasma ($\omega_p / 2\pi = 0.1\,\mathrm{THz}$) with a +z-directed magnetic bias with $\omega_c = 0.5\omega_p$. The effect of loss is modeled by the collision frequency $\Gamma$. The metamaterial slab is periodic along the *x* and *z* directions and has thickness *L* along the *y* direction. We chose a periodic array of short horizontal electric dipoles (oriented along the *x*-direction) as the excitation. The dipole array emits a plane wave that illuminates the metamaterial slab along the normal direction.

Figure 3a (black symbols) shows that similar to the continuum model, the metamaterial structure enables an anomalous wave tunneling at very low frequencies ($\omega = 0.095\omega_p$), where one would expect the electron gas and the wire grid to completely block the wave propagation. The effect holds even in presence of realistic loss values ($\Gamma = 0.1\omega_p$). However, the transmission level with the metamaterial realization is weaker than for the continuum model (see Fig. 2). The matching between the metamaterial slab and the air regions can be noticeably improved by using quarter-wavelength transformers at the input and output interfaces (blue symbols in Fig. 3a). In this case, the transmission amplitude may reach 15% for the thickness $L = 3a$. Figure 3b shows that the transmission level depends significantly on the thickness, due to the excitation of Fabry-Pérot resonances, further supporting that the metamaterial really supports a propagating mode.

Figure 3c shows a time snapshot of the magnetic field at the frequency $\omega = 0.095\omega_p$ and for a thickness $L = 5a$. The time animation of the fields is available in the supplementary information [39] and reveals that in each metal loop the wave follows a rotating motion, such that the energy tends to circulate in closed orbits and the fields have a nontrivial angular momentum [23, 30]. This property can be understood as a consequence of the excitation of topological edge modes (magneto-plasmons) at the wire grid-gyrotropic material interface, due to the different topological nature of the two material phases. The dispersion of the magneto-plasmon mode is depicted in Fig. 1b for the case of a planar interface ($y$=0) between the gyrotropic host material and a PEC. The edge mode dispersion is found as explained in Refs. [9, 20, 21]. As seen, the low-frequency unidirectional edge mode propagates exclusively towards the +$x$-direction in the spectral range $0 < \omega < \omega_c$. When, the gyrotropic material fills a closed metallic cavity the low-frequency edge-mode will go around the cavity walls following an anti-clockwise motion [23], consistent with the winding motion of the magneto-plasmons in each loop of the wire grid. Thus, the anomalous transparency effect can be understood as a consequence of the excitation of topological modes that create coupled vortices of the electromagnetic field (see Fig. 3c and the time animation in Ref. [39]).

Incidentally, some naturally available materials may be as well regarded as a mixture of two inequivalent topological phases with features analogous to those of the model of Eq. (2). For example, let us consider a situation where the microscopic drift current of a plasma is dictated by two species of current carriers (e.g., with different effective mass or different density), i.e., the electron gas is formed by two distinct current "channels". Each species of carriers reacts differently to a bias field, and thereby the combined response of the two-channels may originate the different terms of the dielectric function (2). An illustration of this concept is provided by standard semiconductors, for example GaAs. The dielectric function of GaAs has three contributions: (i) the bound charges, (ii) the free electrons and (iii) the electron-holes. The response of the bound charges is insensitive to a bias magnetic field and may be described by a static permittivity term, $\varepsilon_s = 12.8$. On the other hand, both the electrons and the holes originate drift-currents, yielding a multi-component plasma. The dielectric function of GaAs is of the form:

$$\varepsilon_t = \varepsilon_s - \frac{\omega_{pe}^2}{\omega^2 - \omega_{ce}^2} - \frac{\omega_{ph}^2}{\omega^2 - \omega_{ch}^2}, \qquad \varepsilon_g = \frac{1}{\omega}\frac{\omega_{pe}^2 \omega_{ce}}{\omega_{ce}^2 - \omega^2} + \frac{1}{\omega}\frac{\omega_{ph}^2 \omega_{ch}}{\omega_{ch}^2 - \omega^2}. \qquad (4)$$

with $\omega_{pe}$ and $\omega_{ph}$ ($\omega_{ce}$ and $\omega_{ch}$) the plasma (cyclotron) frequencies for electrons and holes. Interestingly, comparing with Eq. (1), one sees that the dielectric function of the semiconductor is the same as for a mixture of two gyrotropic materials with parameters $\left(\omega_{pe}, \omega_{ce}\right)$ and $\left(\omega_{ph}, \omega_{ch}\right)$. For GaAs the effective masses of electrons and holes are related as $m_e^* = 0.134 m_h^*$ [41, 42]. Hence, the cyclotron frequencies of the two species are linked as $\omega_{ch} = -0.134\omega_{ce}$, and hence have *opposite signs*. Since the sign of the gap Chern number is linked to the sign of the cyclotron frequency [20], the material phases determined by each of the current carriers (electrons or holes) are topologically inequivalent. Indeed, for a +z-directed bias magnetic field, the band structure determined by the gyrotropic response with parameters $\omega_{pe}, \omega_{ce}$ ($\omega_{ph}, \omega_{ch}$) has a low-frequency band-gap with gap Chern number $-1$ ($+1$).

Figure 4a shows the band diagram of GaAs with and without a bias magnetic field. The plasma frequency ($\omega_p = \left(e^2 n / \varepsilon_0 m^*\right)^{1/2}$) depends on the concentration ($n$) and effective mass ($m^*$) of each carrier species. For an intrinsic semiconductor the concentration of electrons and holes is identical, and for the GaAs case $\omega_{ph} = 0.37 \omega_{pe}$. As seen, when $\omega_{ce} = 0 = \omega_{ch}$ (unbiased semiconductor) the dispersion of the modes (black solid line in Fig. 4a; the flat band associated with longitudinal modes is not shown) has a single band-gap near $\omega \approx \omega_{pe}/\sqrt{\varepsilon_s}$. Note that without the bias magnetic field both the electron and hole phases are trivial. In contrast, with a bias magnetic field the electron and hole phases become topologically distinct due to the different sign of the cyclotron frequencies. Hence, similar to the permittivity model (2), the combination of the two distinct topologically phases (which can be observed simply by applying a static bias magnetic field) determines the emergence of a low-frequency transparency window. The transparency window moves to lower frequencies as the bias magnetic field is reduced ($\omega_{ce}/\omega_{pe}$ decreases) as shown in Fig. 4b, analogous to Fig. 1d. It is interesting to note that the "electrons" and "holes" determine two independent current channels, somewhat analogous to the metamaterial design wherein the two material components (metal and magnetized plasma) also determine different paths for the electric current.

In summary, we theoretically demonstrated that by mixing two distinct topological material phases it is possible to create unusual conditions for wave propagation in a spectral range wherein the two phases are impenetrable by light. An electron gas with two current channels (e.g., intrinsic semiconductors) may provide an ideal platform to implement such a structure. In addition, we proposed a realistic metamaterial implementation of the suggested system. Detailed numerical simulations confirm that a material with two distinct topological phases enables, indeed, an anomalous wave tunneling at extremely low frequencies, and opens thus new inroads and opportunities for topological effects in the terahertz and microwave ranges.

**Acknowledgement:** This work is supported in part by Fundação para a Ciência e a Tecnologia with the grants PTDC/EEITEL/4543/2014 and UID/EEA/50008/2013

# References

[1] A. Figotin, I. Vitebsky, "Nonreciprocal magnetic photonic crystals," *Phys. Rev. E* **63**, 066609, (2001).
[2] Z. Wang, Y. D. Chong, J. D. Joannopoulos, M. Soljačić, "Refection-free one-way edge modes in a gyromagnetic photonic crystal", *Phys. Rev. Lett.* **100**, 013905 (2008).
[3] Z. Yu, G. Veronis, Z. Wang, S. Fan, "One-way electromagnetic waveguide formed at the interface between a plasmonic metal under a static magnetic field and a photonic crystal", *Phys. Rev. Lett.* **100**, 023902, (2008).
[4] K. Fang, Z. Yu, V. Liu, S. Fan, "Ultracompact nonreciprocal optical isolator based on guided resonance in a magneto-optical photonic crystal slab", *Opt. Lett.* **36**, 4254-4256 (2011).
[5] D. L. Sounas and C. Caloz, "Gyrotropy and nonreciprocity of graphene for microwave applications," *IEEE Trans. Microw. Theory Tech.*, **60**, 901, (2012).
[6] T. Kodera, D. L. Sounas, C. Caloz, "Artificial Faraday rotation using a ring metamaterial structure without static magnetic field," *Appl. Phys. Lett*. **99**, 031114 (2011).
[7] D. L. Sounas, T. Kodera, C. Caloz, "Electromagnetic modeling of a magnetless and nonreciprocal gyrotropic metasurface", *IEEE Trans. Antennas Propag*. **61**, 221-231 (2013).
[8] Z. Yu, S. Fan, "Complete optical isolation created by indirect interband photonic transitions", *Nat. Photonics* **3**, 91–94 (2009).
[9] A. R. Davoyan and N. Engheta, "Theory of Wave Propagation in Magnetized Near-Zero-Epsilon Metamaterials: Evidence for One-Way Photonic States and Magnetically Switched Transparency and Opacity", *Phys. Rev. Lett*. **111**, 257401, (2013).
[10] A. M. Mahmoud, A. R. Davoyan, N. Engheta, "All-passive nonreciprocal metastructure", *Nat. Communications* **6**, 8359 (2015).

[11] F. R. Prudêncio, M. G. Silveirinha, "Optical isolation of circularly polarized light with a spontaneous magnetoelectric effect", *Phys. Rev. A*, **93**, 043846, (2016).

[12] M. G. Silveirinha, "PTD Symmetry Protected Scattering Anomaly in Optics", *Phys. Rev. B*, **95**, 035153, (2017).

[13] D. L. Sounas, A. Alù, "Non-reciprocal photonics based on time modulation", *Nat. Photon.*, **11**, 774, (2017).

[14] T. A. Morgado, M. G. Silveirinha, "Drift-induced Unidirectional Graphene Plasmons", arXiv:1711.08367.

[15] F. D. M. Haldane, S. Raghu, "Possible realization of directional optical waveguides in photonic crystals with broken time-reversal symmetry", *Phys. Rev. Lett.*, **100**, 013904 (2008).

[16] Z. Wang, Y. Chong, J. D. Joannopoulos and M. Soljačić, "Observation of unidirectional backscattering immune topological electromagnetic states", *Nature*, **461**, 772, (2009).

[17] D. Jin, L. Lu, Z. Wang, C. Fang, J. D. Joannopoulos, M. Soljacic, L. Fu, and N. X. Fang, "Topological magnetoplasmon," *Nat. Commun.* **7**, 13486 (2016).

[18] L. Lu, J. D. Joannopoulos, and M. Soljačić, "Topological photonics", *Nat. Photon.*, **8**, 821 (2014).

[19] T. Ozawa, H. M. Price, A. Amo, N. Goldman, M. Hafezi, L. Lu, M. C. Rechtsman, D. Schuster, J. Simon, O. Zilberberg, I. Carusotto, "Topological Photonics", arXiv:1802.04173, (2018).

[20] M. G. Silveirinha, "Chern Invariants for Continuous Media", *Phys. Rev. B*, **92**, 125153, (2015); M. G. Silveirinha, "Topological classification of Chern-type insulators by means of the photonic Green function", *Phys. Rev. B*, **97**, 115146, (2018).

[21] M. G. Silveirinha, "Bulk edge correspondence for topological photonic continua", *Phys. Rev. B*, **94**, 205105, (2016).

[22] M. G. Silveirinha, "Proof of the bulk-edge correspondence through a link between topological photonics and fluctuation-electrodynamics", arXiv:1804.02190, (2018).

[23] M. G. Silveirinha, "Quantized Angular Momentum in Topological Optical Systems", arXiv:1803.07121, (2018).

[24] M. G. Silveirinha, "A Metamaterial Homogenization Approach with Application to the Characterization of Microstructured Composites with Negative Parameters", *Phys. Rev. B*, **75**, 115104, (2007).

[25] M. Xiao and S. Fan, "Photonic Chern insulator through homogenization of an array of particles", *Phys. Rev. B*, **96**, 100202(R) (2017).

[26] A. H. Sihvola, *Electromagnetic Mixing Formulas and Applications*, IET (London), (1999).

[27] E. D. Palik, R. Kaplan, R. W. Gammon, H. Kaplan, R. F. Wallis, and J. J. Quinn II, "Coupled surface magnetoplasmon-optic-phonon polariton modes on InSb", *Phys. Rev. B*, **13**, 2497, (1976).

[28] E. Moncada-Villa, V. Fernández-Hurtado, F. J. García-Vidal, A. García-Martín and J. C. Cuevas, "Magnetic field control of near-field radiative heat transfer and the realization of highly tunable hyperbolic thermal emitters", *Phys. Rev. B*, **92**, 125418, (2015).

[29] J. A. Bittencourt, *Fundamentals of Plasma Physics*, 3rd Ed. Springer-Verlag, NY, (2010).

[30] M. G. Silveirinha, "Topological Angular Momentum and Radiative Heat Transport in Closed Orbits", *Phys. Rev. B*, **95**, 115103, (2017).

[31] S. A. H. Gangaraj, A. Nemilentsau, G. W. Hanson, "The effects of three-dimensional defects on one-way surface plasmon propagation for photonic topological insulators comprised of continuum media", *Sci. Rep.*, 6, 30055 (2016).

[32] S. A. H. Gangaraj, G. W. Hanson, "Topologically protected unidirectional surface states in biased ferrites: duality and application to directional couplers", *IEEE Antennas Wireless Propag. Lett.*, **16**, 449, (2016).

[33] M. Silveirinha, N. Engheta, "Tunneling of electromagnetic energy through sub-wavelength channels and bends using near-zero-epsilon materials", *Phys. Rev. Lett.*, **97**, 157403, (2006).

[34] M. G. Silveirinha, and N. Engheta, "Theory of Supercoupling, Squeezing Wave Energy, and Field Confinement in Narrow Channels and Tight Bends Using Epsilon-Near-Zero Metamaterials", *Phys. Rev. B*, **76**, 245109, (2007).

[35] N. Engheta, "Pursuing Near-Zero Response", *Science*, **340**, pp. 286-287, (2013).

[36] M. G. Silveirinha, C. A. Fernandes, "Homogenization of 3D- Connected and Non-Connected Wire Metamaterials", *IEEE Trans. on Microwave Theory and Tech.*, **53**, 1418, (2005).

[37] M. G. Silveirinha, C. A. Fernandes, "Transverse Average Field Approach for the Characterization of Thin Metamaterial Slabs", *Phys. Rev. E*, **75**, 036613, (2007).

[38] P.A. Belov, R. Marqués, S. I. Maslovski, I.S. Nefedov, M. Silveirinha, C. R. Simovsky, S. A. Tretyakov, "Strong spatial dispersion in wire media in the very large wavelength limit", *Phys. Rev. B*, **67**, 113103, (2003).

[39] Supplementary online materials with the time animation of Fig. 3c.

[40] CST Microwave Studio 2017 (http://www.cst.com).

[41] M. R. Amin, "Quantum effects on compressional Alfven waves in compensated semiconductors", *Physics of Plasmas*, **22**, 032303, (2015).

[42] A. Moradi, "Surface and bulk plasmons of electron-hole plasma in semiconductor nanowires", *Physics of Plasmas*, **23**, 114503 (2016);.

# Figures

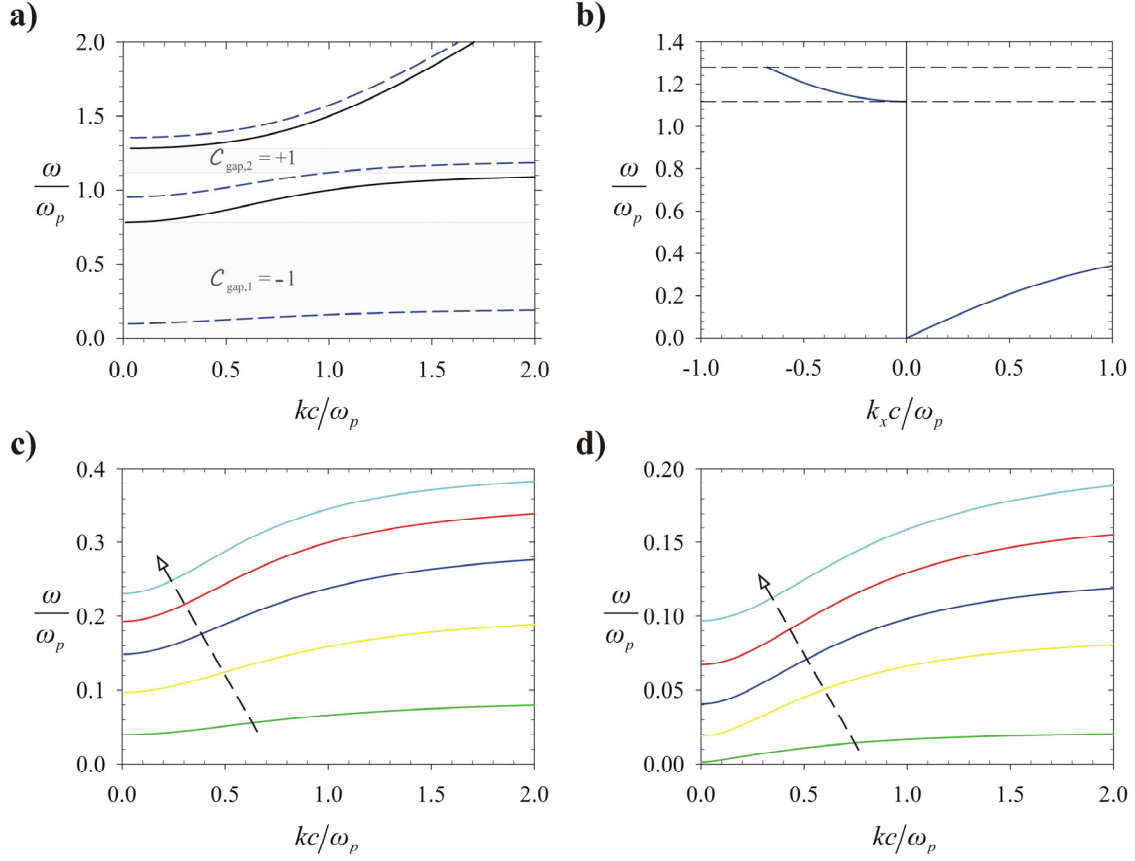

Fig. 1. (**a**) Band diagram of the composite material for propagation in the *xoy*-plane (TM-polarized waves) with $\omega_c = 0.5\omega_p$ for i) $\alpha = 0$ (solid black lines) and ii) $\alpha = 0.5$ (dashed blue lines). (**b**) Dispersion of the topological edge modes supported by a biased plasma ($\omega_c = 0.5\omega_p$ in the region $y > 0$) and a PEC interface ($y = 0$) for propagation along the *x*-direction. (**c**) low frequency band for $\alpha = 0.5$ and $\omega_c/\omega_p = 0.2, 0.5, 0.8, 1.1, 1.4$ (the arrow indicates the direction of increasing $\omega_c/\omega_p$) (**d**) low-frequency band for $\omega_c = 0.5\omega_p$ and $\alpha = 0.05, 0.2, 0.3, 0.4, 0.5$ (the arrow indicates the direction of increasing $\alpha$).

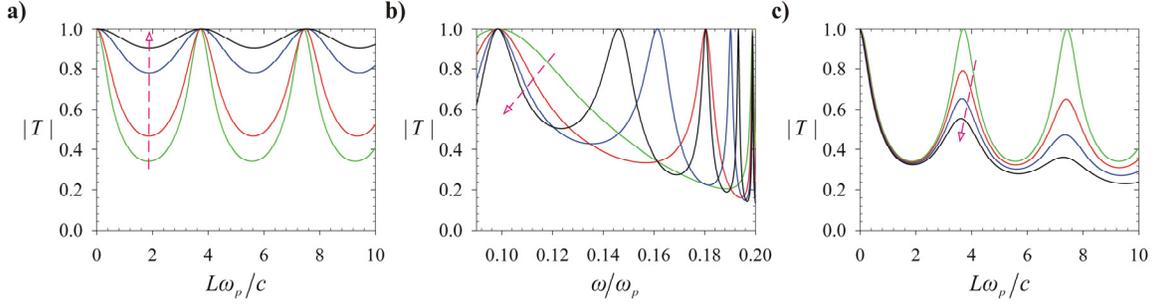

Fig. 2. **(a)** Amplitude of the transmission coefficient as function of the normalized thickness $L\omega_p/c$ at the frequency $\omega = 0.15\omega_p$ for $\omega_c = 0.5\omega_p$ and $\alpha = 0.5$ and for the incidence angles: $\theta^{inc} = 0°, 45°, 70°, 80°$ (the arrow indicates the direction of increasing $\theta^{inc}$) **(b)** Amplitude of the transmission coefficient as function of $\omega/\omega_p$ for $\omega_c = 0.5\omega_p$, $\alpha = 0.5$ and $\theta^{inc} = 0$ and for the normalized thicknesses: $L\omega_p/c = 1.0, 2.0, 3.0, 4.0$ (the arrow indicates the direction of increasing $L$). **(c)** Similar to **(a)** with $\theta^{inc} = 0°$ but for the plasma collision frequency: $\Gamma/\omega_p = 0, 0.005, 0.01, 0.015$ (the arrow indicates the direction of increasing $\Gamma/\omega_p$).

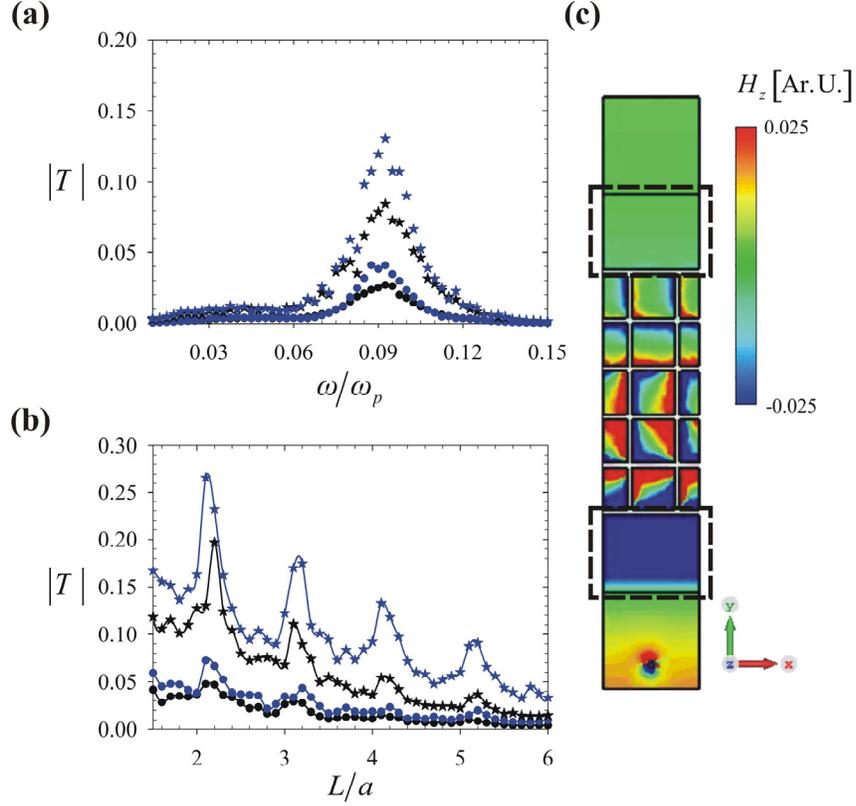

Fig. 3. **(a)** Amplitude of the transmission coefficient as function of the normalized frequency $\omega/\omega_p$ for $\omega_p/(2\pi)=0.1$ THz, $\omega_c=0.5\omega_p$, $a=4c/\omega_p=1.91$ mm, $\theta^{inc}=0$, and $L=3a$ for $\Gamma=0.1\omega_p$ (black symbols) and for $\Gamma=0.05\omega_p$ (blue symbols), without impedance transformers (circle symbols) and with $\lambda/4$ transformers with $\varepsilon_{trans}=6.7$ (star symbols). **(b)** Amplitude of the transmission coefficient as function of the normalized thickness $L/a$ at the frequency $\omega=0.095\omega_p$. The remaining structural parameters are as in (a). **(c)** Time snapshot of the magnetic field ($H_z$) emitted by the dipole array at the frequency $\omega=0.095\omega_p$, for $L=5a$, $\Gamma=0.1\omega_p$, and $\varepsilon_{trans}=6.7$. The remaining structural parameters are same as in panel (a). The black dashed rectangles indicate the location of the $\lambda/4$ transformers. The results are obtained with a full wave simulator [40].

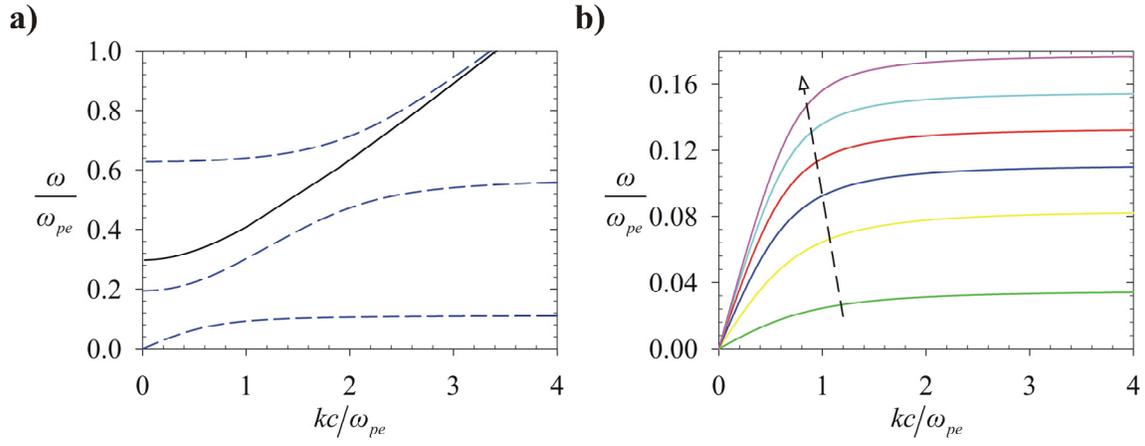

Fig. 4. Band diagram of GaAs for propagation in the *xoy*-plane. **(a)** Unbiased (black solid line) and biased (blue dashed lines) GaAs with $\omega_{ce}/\omega_{pe} = 0.5$. **(b)** the low frequency band for different values of the field bias $\omega_{ce}/\omega_{pe} = 0.1, 0.3, 0.5, 0.7, 0.9, 1.1$ (the arrow indicates the direction of increasing $\omega_{ce}/\omega_{pe}$)